\documentclass[12pt,leqno]{article}
\newcommand{\comment}[1]{}
\usepackage{amssymb,amsmath}
\usepackage{graphicx,psfrag,color}

\definecolor{cy}{rgb}{0,.5,.5}
\definecolor{rd}{rgb}{.7,0,0}
\definecolor{mg}{rgb}{.6,0,.6}
\definecolor{bl}{rgb}{0,0,1}
\definecolor{gr}{rgb}{0,.5,0}

\newcommand{\rd}{\textcolor{rd}}   %
\newcommand{\cy}{\textcolor{cy}}   %
\newcommand{\mg}{\textcolor{mg}}
\newcommand{\bl}{\textcolor{bl}}   %
\newcommand{\rr}{\mbox{\bf r}}
\newcommand{\gr}{\textcolor{gr}}

\begin{document}

\title{A twist in the geometry of rotating black holes:\\
seeking the cause of acausality}
\author{Hajnal Andr\'eka, Istv\'an N\'emeti and Christian W\"uthrich}
\date{5 November 2007}
\maketitle

\begin{abstract}\noindent
We investigate Kerr-Newman black holes in which a rotating charged ring-shaped singularity induces a region which contains CTCs. Contrary to popular belief, it turns out that the time orientation of the CTC is {\em opposite} to the direction in which the singularity or the ergosphere rotates. In this sense, CTCs ``counter-rotate'' against the rotating black hole. We have similar results for all spacetimes sufficiently familiar to us in which rotation induces CTCs. This motivates our conjecture that perhaps this counter-rotation is not an accidental oddity particular to Kerr-Newman spacetimes, but instead there may be a general and intuitively comprehensible reason for this.
\end{abstract}

\section{Introduction}
In the present note we investigate rotating black holes and other
generally relativistic spacetimes where rotation of matter might
induce closed timelike curves (CTCs), thus allowing for a ``time
traveler'' who might take advantage of this spacetime structure.
Most prominently, we will discuss Kerr-Newman black holes in which a
rotating charged ring-shaped singularity induces a region which
contains CTCs. Due to the electric charge of
the singularity, this region is not confined to within the analytic
extension ``beyond the singular ring'', but extends into the side of
the ring-singularity facing the asymptotically flat region, from
whence the daring time traveler presumably embarks upon her
journey. Interestingly, some kind of ``counter-rotational
phenomenon'' occurs here. If a potential time traveler wants to use
our Kerr-Newman black hole for traveling into her past, she will
have to orbit along the CTC in the direction opposite to that of the
rotation of the black hole. In technical words, the time orientation
of the CTC is opposite to the direction in which the singularity or
the ergosphere rotates.

This state of affairs is at odds with the qualitative, intuitive
explanations for the mechanism creating CTCs presented in most
popular books. This will motivate the questions
formulated in section~\ref{expl-sec}.%
\footnote{References to popular books offering such intuitive, but
misleading, explanations will also be given in
section~\ref{expl-sec}.} We have similar results for all spacetimes
sufficiently familiar to us in which rotation induces CTCs. This
motivates our conjecture that perhaps this counter-rotation is not
an accidental oddity particular to Kerr-Newman spacetimes, but
instead there may be a general and intuitively comprehensible reason
for this.

Understanding what we take to be the most promising candidate
mechanism to produce CTCs in an otherwise causally well-behaved
spacetime, i.e.\ the counter-rotational phenomenon mentioned above,
is of paramount importance in evaluating the causal stability of
generally relativistic spacetimes. Since one possibility to violate
Hawking's chronology protection conjecture or, more generally, the
strong form of Penrose's cosmic censorship conjecture is through the
emergence of acausal features via such a mechanism \cite{EW}, the
present paper contributes to efforts directed at the larger projects
of understanding chronology protection and cosmic censorship in
general relativity. Furthermore, the issues discussed here are also
motivated by discussions in our works \cite{EN02} and \cite{ND}. In
those papers it turned out that studying the geometry of rotating
black holes can be relevant to some far-reaching considerations
in the foundation of mathematics and logic. The counter-rotational phenomenon in Kerr-Newman
spacetime was already noted explicitly in
\cite[p.~55]{Wdis}, albeit without further analysis.
A fascinating book providing a broad perspective for the
presently discussed matters is Earman~\cite{Bangs}.
\bigskip

\section{A counter-rotational phenomenon in Kerr spacetime}\label{kerr-sec}
\begin{figure}
\begin{center}
\psfrag*{text1}[b][b]{\bl{$\theta=0$, axis of rotation}}
\psfrag*{r0}[b][b][0.8]{\,\mg{$r=0$}} \psfrag*{r-1}[b][b][0.8]{
\mg{$r=-1$}} \psfrag*{r1}[b][b][0.8]{\mg{$r=1$}}
\psfrag*{ri}[b][b][0.8]{\mg{$r=-\infty$}}
\psfrag*{theta}[l][l]{\bl{$\theta$}}
\psfrag*{text2}[l][l]{\bl{$\theta=\pi/2$}}
\psfrag*{text3}[l][l][0.8]{\mg{$r\in(-\infty,0]$}}
\psfrag*{text4}[l][l][0.8]{\mg{$r\in[0,\infty)$}}
\includegraphics[keepaspectratio, width=0.7\textwidth]{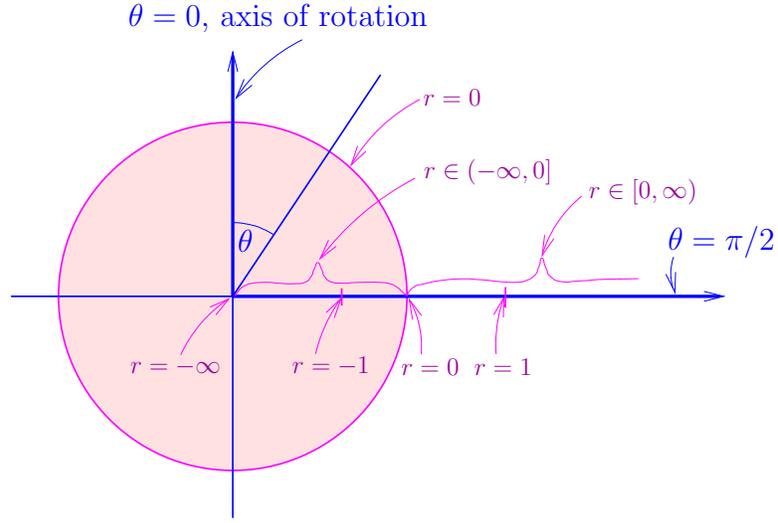}
\end{center}
\caption{\label{coordinates} Illustrating coordinates $r$ and
$\theta$. Radius $r$ is drawn as $e^r$, so that the shaded circle
covers the negative values for $r$.}
\end{figure}

We use the standard (Boyer-Lindquist) coordinates
$t,r,\varphi,\theta$ for Kerr spacetime which appear e.g.\ in
Hawking-Ellis \cite[p.~161]{HE}, O'Neill \cite[pp.~57-59]{ON95},
Wald \cite[p.~313]{Wald84}, Misner-Thorne-Wheeler \cite[p.~877, item
(33.2)]{MTW}. Of these four coordinates, $t,r$ range over the reals,
i.e.\ $-\infty<t,r<\infty$ while $\varphi,\theta$ are spherical
coordinates. In pictures, the radius $r$ is drawn as $e^r$, so
$r=-\infty$ is at the center of the figure and the $r$ coordinate is
negative within the sphere indicated in the drawing, see Figure
\ref{coordinates}.

Using these coordinates, the metric tensor field $g$ of the Kerr
spacetime is given by
\smallskip

\begin{description}
\item
$g_{tt} = -1 + 2Mr/\rho^2$, \quad where $\rho^2=r^2+a^2\cos^2\theta$,
\item
$g_{rr} = \rho^2/\Delta$, \quad where $\Delta=r^2-2Mr+a^2$,
\item
$g_{\theta\theta} = \rho^2$,
\item
$g_{\varphi\varphi} =
(r^2+a^2+[2Mra^2\sin^2\theta]/\rho^2)\sin^2\theta$,
\item
$g_{t\varphi} = -2Mra\sin^2\theta/\rho^2$ \quad and all the other
$g_{ij}$'s are zero.
\end{description}
\smallskip
Here $a$ denotes the angular momentum per unit mass of the rotating ring, while $M$
is called its mass, cf.\ e.g.\ \cite[p.~58]{ON95}.

We will concentrate on the so-called ``slow-Kerr" case when
$0<a^2<M^2$. In this case there are two event horizons defined
by the roots of $\Delta=0$:
$$
r_{\pm} = M \pm \sqrt{M^2-a^2}.
$$
In case the Kerr black hole spins sufficiently fast ($a^2>M^2$),
these event horizons vanish and we would be faced with a naked
singularity. Since such beasts may be regarded as
unphysical, we disregard the ``fast-Kerr'' case.%
\footnote{However, everything in this paper applies to the fast-Kerr
case, too, except that some formulations would need to be adapted in
order to equally apply to the fast-Kerr case.}
 We will be
interested in the ``innermost" region of the black hole, defined by
$r<r_-$, which is where the CTCs (i.e.\ closed timelike curves) are.
This part of the spacetime is called {\em block III} or {\em
negative exterior Kerr spacetime (EKN$-$)}. The so-called
``equatorial plane" is defined by $\theta=\pi/2$. This plane
contains, in block III, the so-called ring-singularity
$$\Sigma = \{ \langle t,r,\varphi,\theta\rangle :
r=0\mbox{ and }\theta=\pi/2\}.$$ For the Kerr case, i.e.\ for an
{\em uncharged} rotating black hole, the CTCs transpire inside and
close to this ring-singularity, i.e.\ CTCs are found in regions
where $r$ is negative but has small absolute value. This part of the
spacetime belongs to what \cite{ON95} calls the {\em Time Machine}.%
\footnote{The notion of a time machine, which has previously been
used rather loosely in the physics literature, has recently been
subjected to a more rigorous analysis in \cite{takearide} and
\cite{EW}. In the full understanding of the vagueness of O'Neill's
terminology, we stick to it for simplicity of discussion, as our
main issue here is not the question of what should qualify as a time
machine.} On the other hand, the part outside the ring-singularity
and sufficiently close to it belongs to the so-called
``ergosphere.'' The {\em ergosphere} (in block III) is defined to be
the place where the vectors $\partial_t$ parallel to the
``time-axis" are not timelike but spacelike. An ergosphere, thus, is
a region of spacetime where no observer can remain still with
respect to the coordinate system in question. For static
black holes, such as those described by Schwarzschild spacetime, the
outer limit of the ergosphere coincides with the black hole's event
horizon. This is no longer the case when the black holes revolve.
Then, the faster the black hole rotates, the more the ergosphere
grows beyond the outer horizon $r=r_+$. The $(r,\varphi)$-surfaces
close to the equatorial plane, i.e.\ the surfaces with fixed $t$ and
fixed small $\cos\theta$, look similar, except that they contain no
singularity (if $\cos\theta\ne 0$).

All our investigations below are strongly connected to the {\em
time-orienta\-tion} of the spacetime being discussed. Therefore, we recall for
convenience that in O'Neill's book \cite[p.76 lines 7-8 and p.60
Def.2.1.2.]{ON95} the time-orientation for block III of Kerr
spacetime is defined by the vector-field \label{orient}
$$
V(p) :=
(a^2+r^2)\partial_t(p) + a\partial_{\varphi}(p).
$$
With the metric and a time-orientation at hand, the notion of a
well-pa\-ram\-e\-ter\-ized future-pointing curve makes sense now.

Assume that the lifeline of a particle $\alpha$ is given by the
well-parameterized, future-pointing curve $\alpha(\tau)$ with
$\tau\in I$ where $I$ is an interval of the reals. We say that
$\alpha$ is {\em rotating in the direction $\partial_\varphi$}
(at $\tau_0\in  I$) if with increasing proper time $\tau$ the
values of the $\varphi$-component are increasing (at
$\tau_0$), i.e.\ if ${d}\alpha_\varphi(\tau)/{d}\tau$ is
positive (at $\tau_0$). Sloppily, we could write this as
``$d\varphi/d\tau>0$" on $\alpha$ (at $\tau_0$). We say that the direction of rotation of
$\alpha$ (at $\tau_0$) is $-\partial_\varphi$ if the latter value is
negative.

\begin{figure}[!h]
\begin{center}\bigskip
\psfrag*{dt}[b][b]{\bl{$\partial_t$}}
\psfrag*{df}[b][b]{\bl{$\partial_\varphi$}}
\psfrag*{v}[b][b]{\rd{$V$}} \psfrag*{text1}[l][l]{\it
\label{fig:light1} Light cones in the ergosphere}
\psfrag*{text2}[l][l]{\it Light cones on the CTCs}
\includegraphics[keepaspectratio, width=0.8\textwidth]{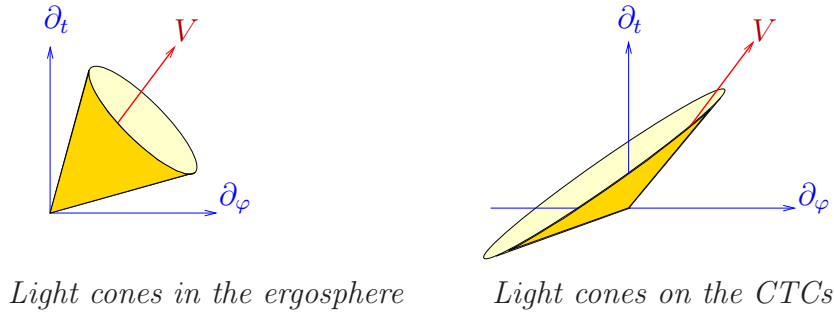}
\end{center}
\caption{\label{fig:light2} Light cones in the ergosphere and light
cones on the CTCs look different.}
\end{figure}

It is known that any particle in the ergosphere must rotate in the
$\partial_\varphi$ direction, it is not possible to avoid rotating
or to rotate in the $-\partial_\varphi$ direction in the ergosphere
(see e.g.\ \cite[Lemma 2.4.4]{ON95}). The reason is that in the
ergosphere the light cones are tilted in the $\varphi$ direction,
i.e.\ the light cones in the $(t,\varphi)$-cylinders look like those
in the left-hand side of Figure~\ref{fig:light2}.

The situation is drastically different in the ``Time Machine''
region! There the light cones look like those in the right-hand side
of Figure~\ref{fig:light2}. This means that here it is possible to
orbit in the $-\partial_\varphi$ direction as well as in the
positive $\partial_\varphi$ direction, but any time-traveler (i.e.\
one with $dt/d\tau\le 0$) has to rotate in the $-\partial_\varphi$
direction, for it is the only possibility to construct a path with
$dt/d\tau<0$ (cf.\ Figure \ref{fig:light2}).

Our above statements can be formulated as saying that the ergosphere
and the ``Time Machine'' rotate in opposite directions. By this we
mean that a traveler in the ergosphere and a traveler moving forward
in time in the Time Machine but just preparing for entering a CTC,
i.e.\ with $dt/d\tau > 0$ approaching ($dt/d\tau=0)$, move in
opposite directions. The one in the ergosphere co-moves with the
singularity (or the source) while the one on the CTC, or almost on
the CTC, moves in the opposite direction. This results from
considering the singularity as the source of the field and the
assumption that it rotates in the positive $\varphi$ direction
because the total angular momentum $J$ of the Kerr spacetime with
$a>0$ is positive.\footnote{For details, cf.\
\cite[pp.296-297,314]{Wald84}, \cite[pp.58,179]{ON95}.}

This is just what we dub a {\em counter-rotating effect}: the
property that the CTCs counter-rotate with the ring-singularity (and
with the ergosphere) in the sense just described. In other words,
the time-orientations of the CTCs point in the $-\partial_{\varphi}$
direction while the ergosphere rotates in the opposite,
$+\partial_{\varphi}$ direction. We note that this
counter-rotational effect remains valid if we extend our attention
beyond the equatorial plane defined by $\theta=\pi/2$. We also
note that this counter-rotational effect does not depend on which of
the two possible time-orientations we choose.

We can formulate this counter-rotational effect in a
coordinate-independent way by saying that where the (invariantly
defined) Killing vector field $\partial_{\varphi}$ is timelike, its
time-orientation is negative (and hence a would-be time traveler
must orbit in the $-\partial_{\varphi}$ direction). Here we assume
that the time-orientation is chosen such that the rotation of the
source points in the positive $\varphi$ direction. (We note that our
claim that the CTCs and the ergosphere rotate in the opposite
directions can also be formulated in a coordinate-independent way.)

\section{Counter-rotation in Kerr-Newman spacetime}\label{kn-sec}

At this point one might be tempted to think that perhaps the cause
of the above counter-rotational phenomenon might be found in the
fact that the co-rotating area (ergosphere) and the counter-rotating
area (Time Machine) are separated by the ring-singularity. So
perhaps the counter-rotation can be explained by saying that the
ring-singularity acts like a mirror turning directions into their
negatives. Such a symmetry about the ring-singularity would still
not account for the rotational sense of the ring-singularity itself;
in particular it would keep us wondering why the singularity
co-rotates with the ergosphere but counter-rotates with the CTCs.
The simplest way of seeing that such mirroring about the singular
region cannot possibly give a hint for the diametric revolutions of
the ergosphere and the Time Machine is by looking at Kerr-Newman
spacetimes where the co-rotational and counter-rotational areas are
no longer separated by the ring-singularity.%
\footnote{For completeness we note that a further possibility for
seeing this is staying with Kerr spacetime and focusing attention to
$(r,\varphi)$ hypersurfaces with fixed, small values of $\theta$. In
such a hypersurface the co-rotating and counter-rotating regions
still exist but they are no longer separated by a singularity.}

Kerr-Newman spacetimes describe black holes with angular momentum as
in the Kerr case, but with an electric charge in addition. We
are using the definition of Kerr-Newman spacetime as given e.g.\ in
Misner-Thorne-Wheeler~\cite[pp.~877-881, e.g.\ item (33.2) on
p.~877]{MTW}, Wald~\cite[p.~313, item (12.3.1)]{Wald84},
d'Inverno~\cite[p.~264, item (19.72)]{Dinv}, or
W\"uthrich~\cite{Wdis}.

The Kerr-Newman metric can be obtained from the Kerr metric (in
Boyer-Lindquist coordinates) by simply replacing all occurrences of $2Mr$
with $(2Mr-e^2)$, where $e$ is the electric charge of the black hole.%
\footnote{This statement is a refinement of what \cite{ON95} writes
on p.~61 about the connection between the two metrics.} Thus, the
metric will take on the form
\smallskip
\goodbreak

\begin{description}
\item
$g_{tt} = -1 + (2Mr-e^2)/\rho^2$, \quad where $\rho^2=r^2+a^2\cos^2\theta$,
\item
$g_{rr} = \rho^2/\Delta$, \quad where $\Delta=r^2-2Mr+e^2+a^2$,
\item
$g_{\theta\theta} = \rho^2$,
\item
$g_{\varphi\varphi} =
(r^2+a^2+[(2Mr-e^2)a^2\sin^2\theta]/\rho^2)\sin^2\theta$,
\item
$g_{t\varphi} = (e^2-2Mr)a\sin^2\theta/\rho^2$ \quad and all the other
$g_{ij}$'s are zero.
\end{description}
\smallskip

\begin{figure}
\begin{center} \bigskip
\psfrag*{t}[b][b]{\bl{$t$}}
\psfrag*{timeorientation}[b][b]{\shortstack[l]{\mg{time-orientation}\\
\mg{of CTC\,!}}}
\psfrag*{ringsingularity}[b][b]{\shortstack[l]{rotation of\\
ring-singularity\\ ($r=0$)}} \psfrag*{ctc}[l][l]{\shortstack[l]{\mg{CTC at}\\
\mg{$r\le r_1<r_0$}}} \psfrag*{dt}[r][r]{\bl{$\partial_t$}}
\psfrag*{df}[r][r]{\rd{$\partial_\varphi$}}
\psfrag*{eventhorisons}[l][l]{\bl{event horizons}}
\psfrag*{erect}[l][l]{\shortstack{light cones are erect\\ at $r_0$}}
\psfrag*{rm}[l][l]{\bl{$r_-$}} \psfrag*{rp}[l][l]{\bl{$r_+$}}
\psfrag*{ergosphere}[l][l]{\cy{ergosphere}}
\psfrag*{r}[l][l]{\bl{$r$}}
\includegraphics[keepaspectratio, width=\textwidth]{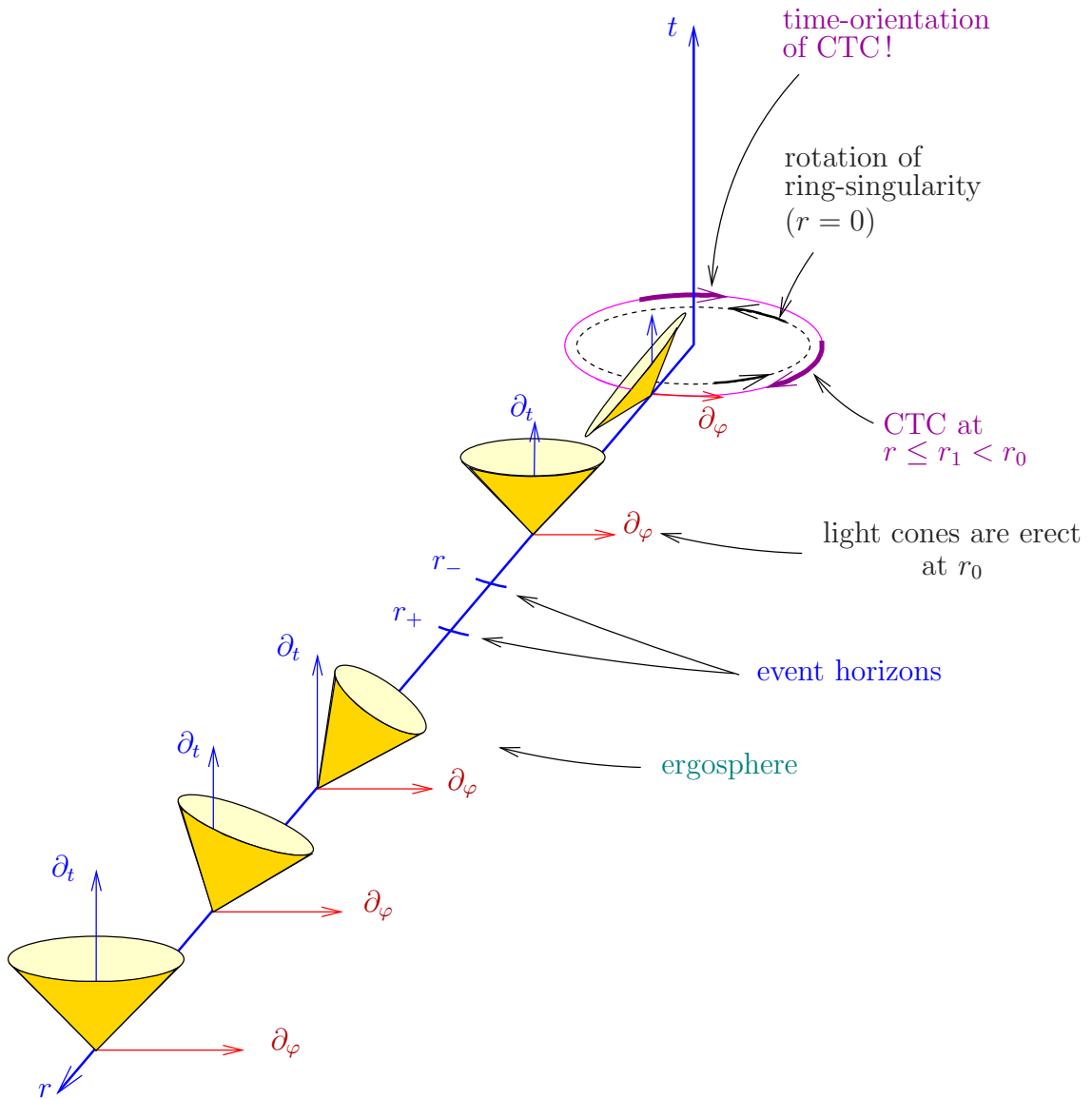}
\end{center}
\caption{\label{series-fig}This is how light cones behave in
Kerr-Newman spacetimes. For $r_0, r_1$ and computations see
section~\ref{app-sec}, especially Figure~\ref{functions-fig}.}
\end{figure}

The vector field $V(p)$ as defined in section \ref{kerr-sec} is an
admissible time-orientation for block III of the Kerr-Newman
spacetime, too. As the focus of the present article is precisely on
block III, this reassurance suffices for our purposes. The
counter-rotating phe\-nomenon described in the previous section
holds in the case of charged Kerr-Newman holes, too. However, this
counter-rotation effect is more poignant than in the Kerr case, for
the following reason. In the Kerr-Newman spacetime there are CTCs
both at small positive values of $r$ (say, at $r=0.1M$) and also at
small negative values of $r$ (say, at $r=-0.1M$) in the equatorial
plane. Thus, the ergosphere and the ``Time Machine" are no longer
separated by the ring-singularity in this case. However, here, too,
the time-orientation of the CTC at $r=0.1M$ points in the direction
$-\partial_{\varphi}$, i.e.\ in the negative $\varphi$ direction.
Using our terminology introduced above, this means that a
``time-traveler'' inhabiting the CTC at $r=0.1M$ orbits in the
direction opposite to the rotation of the massive ring. In other
words, the CTC and the ring-singularity ``rotate'' in opposite
directions just as they did in the Kerr case, as shown in
Figure~\ref{series-fig}.

Moreover, {\em all} CTCs in Kerr and Kerr-Newman spacetimes
counter-rotate with the massive ring. In particular, any ring-shaped
CTC like ${\mathcal R}(\rr)=\{\langle 0,r,\varphi,\theta\rangle :
\theta=\pi/2\mbox{ and }r=\rr\}$ for fixed $\rr$ has
time-orientation $-\partial_{\varphi}$ (for all possible choices of
$\rr$ making ${\mathcal R}(\rr)$ into a CTC).\footnote{We note that
W\"uthrich~\cite[Chapter 6, pp.~75-85]{Wdis} is highly relevant to
our investigations here (e.g.\ it is proved there that in order to
remain on a CTC, in Kerr-Newman spacetime, the ``time-traveler" has
to use a prohibitive amount of fuel). Furthermore, Figures 4.2 and
4.4 on pages 49 and 52 in \cite{Wdis} depict the causality
violating regions for the Kerr and the Kerr-Newman case
respectively.}

\section{Seeking explanations}\label{expl-sec}

In the literature, the apparently {\em standard} account for why and
how rotating matter induces CTCs goes as follows.\footnote{Cf.\
e.g.\ Gribbin~\cite[pp.~145-152, and Fig.~8.1 on p.~151]{Grib83} or
Gribbin~\cite[pp.~214-220 (e.g.\ Figures 7.5 and 7.6 on p.~215 and
p.~218 resp.)]{Grib92}. \cite{Grib92} writes on the bottom of page
217 that ``[t]he critical stage for the light cone tipping, as far
as time travel is concerned, is when the cone is tipped by more than
45 degrees. Since the half-angle of the cone is 45$^o$, ...".
Further references for the standard explanation are e.g.\ Paul
Davies~\cite[first 3 pages of Section 2]{Davies}, Nick
Herbert~\cite[Fig.6-2 on p.105]{NH88}, Paul Horwich \cite[p.~113,
Figure 28]{Hor}, and Clifford A.\ Pickover~\cite[Figure 14.2
(p.~185), Fig.17.1 on p.~224 might also be relevant]{Pick98}.}
Given, for example, a Kerr-Newman black hole, let us mentally move
toward the singularity starting from a point far away from the black
hole. Let us further assume that this vantage point is situated on
the equatorial plane of the black hole. The Kerr-Newman spacetime is
asymptotically flat. Thus, at this distant point, the black hole
does not influence the light cones significantly and we can safely
assume that they are straight or ``vertical" like in Minkowski
spacetime. Let us then move slowly towards the black hole. As it
gets closer, the so-called dragging of
inertial frames caused by the rotation of the black hole comes into
action. This dragging effect is explained in relativity
textbooks via Mach's principle, e.g.\ \cite[pp.319,187,89]{Wald84}
or \cite[pp.547,879,1117-1120]{MTW}. So as an influence of the
rotation of our massive ring (the singularity), the light cones get
tilted in the direction $\varphi$ of the rotation of the ring.
Naturally, the closer we get to the rotating ring, the stronger this
dragging effect, and thus the tilting, will be. The standard
account goes on by saying that eventually the light cones tip over
completely (i.e.\ they dip below the equatorial plane), so they
become approximately ``horizontal," making $\partial_\varphi$
timelike (and $\partial_t$ spacelike). This clearly leads to CTCs by
$\partial_\varphi$ being timelike. This literature thus creates the
false impression that the proper time of the CTCs {\em co-rotates},
rather than {\em counter-rotates}, with the matter content of the
universe. The time orientation of the CTCs, according to this
``official story,'' therefore agrees with the direction of rotation
of the source of the field.

It is important to stress that according to this official story the
time orientation of the so obtained CTCs agrees with the direction $\varphi$
of rotation of the black hole---contrary to what has been established in
sections \ref{kerr-sec} and \ref{kn-sec}. Hence a time traveler using such
CTCs would orbit in the positive $\varphi$ direction, i.e.\ would rotate in
the same direction as the black hole does. If this explanation worked, it would
yield an intuitively convincing, natural explanation for why and how the basic
principles of general relativity lead in certain situations (such as when appropriately
distributed rotating masses are present) to CTCs. One of our main points is that the
above explanation does not work, simply because light cones behave differently in
the relevant spacetimes, as is also illustrated in Figure~\ref{series-fig}.

Let us see how the corrected story goes based on detailed
computations in Kerr-Newman spacetime.  These computations will be
presented in section~\ref{app-sec}. First of all, we emphasize that
the new, corrected story suggested in the present paper does not
offer any kind of {\em explanation} for the creation of CTCs.
Instead, it merely offers a description of the behavior of the light
cones. We maintain, however, that a proper understanding of this
behavior, and particularly the counter-rotation that we are
interested in, at least constitutes an important and promising first
step toward such an explanation. For simplicity we assume that the
charge and rotation of the black hole are sufficient for ensuring
that there are CTCs at positive values of the radius $r$, i.e.\
outside the ring-singularity. Surely enough, the corrected story
begins exactly like the official party line: Distant light cones are
erect, i.e.\ vertical and they start to tilt in the $\varphi$
direction as we begin moving towards the black hole. But this
tilting effect does not grow beyond any limit as we move towards the
ring-singularity. As we move inward (towards the ring), the tilting
grows for a while but then it stops growing, and eventually at a
radius $r_0$ safely outside the time-travel region, the light cone
(in the $t\varphi$-plane) is erect again. From $r_0$ inward, the
tilting is in the other direction, i.e.\ in the $-\varphi$
direction. (Until now, i.e.\ at values greater than $r_0$, the
tilting was always in the $+\varphi$ direction.) Thus, tilting in
the $\varphi$ direction did not result in CTCs, because $r_0$ is
safely outside the time-travel region. Moreover, tilting alone in
the $-\varphi$ direction does not lead to CTCs, either, because from
$r_0$ inward, the time-axis is always inside the light cone (i.e.\
$\partial_t$ is always timelike).

So what happens if we go closer and closer to the ring-singularity?
In other words, what creates the CTCs if not the cone tipping? The
answer is that, eventually, the {\em light cones open up} in the
negative $\varphi$ direction. This second effect is carried to such
an extreme that $-\partial_\varphi$ becomes contained in the future
light cone, for sufficiently small but positive values of the radius
$r$. If we are close enough to the ring-singularity, then time
travel becomes possible---but only if we orbit in the negative
$\varphi$ direction. Hence the time traveler has to orbit in the
direction opposite to the rotation of the black hole. It is this
fact that we refer to as the phenomenon of counter-rotating. Let us
summarize the new story's two most important features. These are:

(i) counter-rotation (i.e.\ the time orientation of the CTCs is
$-\partial_\varphi$), and

(ii) it is not the tilting of the light cones due to the
dragging of inertial frames which leads to
the formation of CTCs, but a second effect, primarily resulting in
their opening up in the negative $\varphi$ direction.

The open issue we would like to raise is the following. The official
story as presented at the beginning of this section provided an
intuitive physical account as to what causes the tipping of the
light cones. To repeat, our computations show that this official
story is not true and that it has to be replaced with the new story
outlined above. Unfortunately, however, we cannot offer an equally
intuitive and suggestive explanation for the new story. Hence, we
would like to issue a challenge to our readers in form of the
following question:

\bigskip
\noindent {\bf Question 1.} Is there a qualitative---and similarly
compelling---explanation as to why the time traveler has to
``counter-rotate" against the rotating ring-singularity in
Kerr-Newman spacetime? Can one find a physical mechanism which
qualitatively explains why and how CTCs are ``created" by rotation
of matter?\bigskip

In the above description of the ``new story" we concentrated on
Kerr-Newman black holes for simplicity. The situation is similar in
many spacetimes where rotation of relatively large masses leads to
the formation of CTCs. For example, our corrected story applies
equally to Kerr black holes, with appropriate modifications to adapt
the train of thought to the fact that in the Kerr case the CTCs only
transpire in the negative radius region.\footnote{To see the
counter-rotation effect in action in the Kerr case, our metaphorical
spaceship must approach the causality-violating region somewhat
``from above'' the equatorial plane, for it would crash into the
ring-singularity otherwise. In other words, we choose some fixed
$\theta$ with $\cos\theta>0$ which, however, must be sufficiently
small for intersecting the causality violating region. This slight
change of itinerary does not alter in principle the
counter-rotational effect we discussed at length for the Kerr-Newman
case.} \footnote{Despite the seemingly widely held presupposition
that the CTCs co-rotate rather than counter-rotate with the black
hole, we seem to have an ally in Brandon Carter, at least as far as
the Kerr case is concerned. He seems to concur with our conclusion
that the outlined counter-rotation does not arise from a coordinate
artefact, but constitutes a tangible physical process when he writes
that ``in order to make up literally for the lost time, the path
must enter the region [where CTCs transpire]. Here time can be
gained but only at the expense of clocking up a large change ({\em
negative} for $a>0$) in the angle $\hat{\varphi}$.'' \cite[p.~1566,
below item (28); our emphasis]{BC68}. A second ally is Robert L.\
Forward~\cite{For} who on p.~172 describes time travel via Kerr
spacetime as ``travelling near the rotating ring in the direction
against rotation of the ring for a number of rotations''. Cf.\ also
his Fig.~10 on p.~176.}

\section{Computations.} \label{app-sec} 
\begin{figure}[!h]
\begin{center} \bigskip
\psfrag*{t}[b][b]{\bl{$t$}} \psfrag*{ergo}[b][b]{\cy{ergosphere}}
\psfrag{gtt}{\cy{$g_{tt}$}}
\psfrag{gff}[c][l]{\mg{$g_{\varphi\varphi}$}}
\psfrag{gtf}{{$g_{t\varphi}$}}
\psfrag*{time}[b][b]{\shortstack[l]{time-\\
machine\\ ($r\le r_1$)}}
\psfrag*{Rp}[r][r]{\cy{$R_+$}} \psfrag*{Rm}[r][r]{\cy{$R_-$}}
\psfrag*{rp}[b][b]{} \psfrag*{rm}[b][b]{} \psfrag*{M}[b][b]{$M$}
\psfrag*{0}[c][l]{\bl{$0$}} \psfrag*{r0}[r][r]{{$r_0$}}
\psfrag*{r1}[r][r]{\mg{$r_1$}} \psfrag*{ctc}[c][l]{{\mg{CTCs}}}
\psfrag*{r}[l][l]{\bl{$r$}}
\includegraphics[keepaspectratio, width=\textwidth]{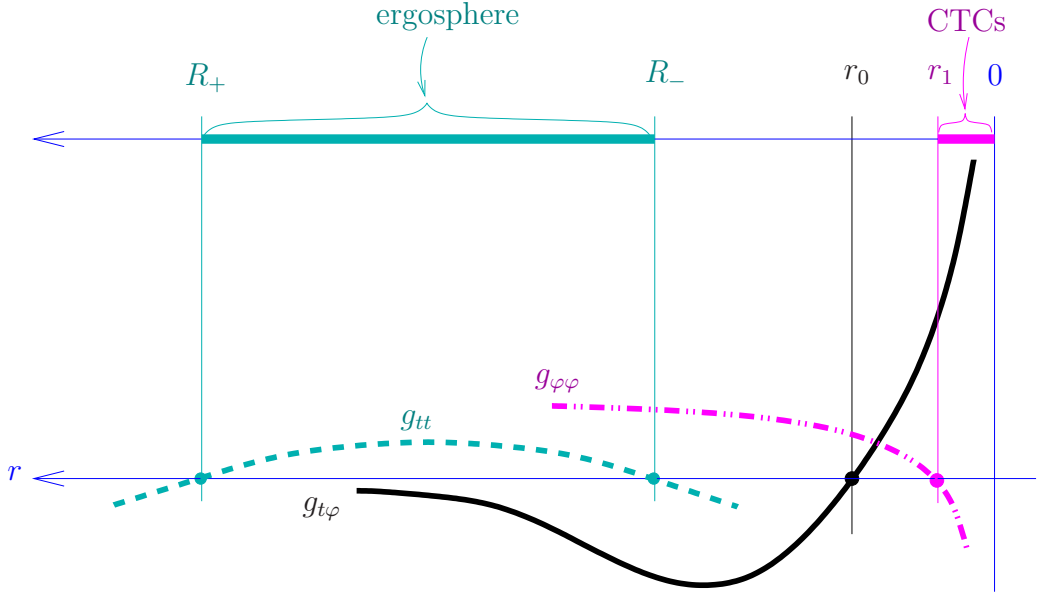}
\end{center}
\caption{\label{functions-fig} Illustration for the Kerr-Newman
metric. The picture faithfully represents the order of the roots
(i.e.\ zeros) of $g_{t\varphi}, g_{tt}$ and $g_{\varphi\varphi}$ and
where these three functions are negative or positive; the picture
does not faithfully represent the proportions of the distances and
their magnitudes. The ergosphere and the ``Time
Machine" (where the CTCs are) are disjoint, and the light cone
is erect between them at $r_0$. In
some sense, $g_{t\varphi}$ represents the tilting of the light
cones. See also Figure~\ref{series-fig}.}
\end{figure}

In this section, we
present computations supporting the claims made in
section~\ref{expl-sec}. We do the computations for the Kerr-Newman
metric; we assume $a\ne 0, e\ne 0$ and $M>\sqrt{a^2+e^2}$. We are in the
equatorial plane, i.e.\ $\theta=\pi/2$, $\cos\theta=0$, and
$\sin\theta=1$. We are interested in what the light cones look like
in the $t\varphi$-planes, as a function of the radius $r$, and we
are interested in positive $r$ only as this suffices to study the
tilting and widening of the light cones. Using the shorthand
\begin{description}
\item[]
$E=(2Mr-e^2)/r^2$\qquad we have
\item[]
$g_{tt}=E-1$,\quad $g_{t\varphi}=-aE$,\quad and\quad
$g_{\varphi\varphi}=r^2+a^2+a^2E$.
\end{description}
These functions are depicted in Figure~\ref{functions-fig}.

Let us first check where the CTCs are. We have CTCs where
$g_{\varphi\varphi}$ is negative, and this is exactly where
\begin{description}
\item[]
$G(r):=r^2g_{\varphi\varphi}=r^4+a^2r^2+2a^2Mr-a^2e^2$
\end{description}
is negative. The radial derivative of $G$ is $G'=4r^3+2a^2r+2a^2M$,
this is everywhere positive because $r\ge 0$ and $M>0$. Hence $G$ is
monotonically increasing (as we move from $0$ to infinity). We have
$G(0)=-a^2e^2<0$ and $G$ tends to infinity as $r$ tends to positive
infinity; hence there is a unique value $r_1$ where $G$ is zero,
above which $G$ is positive and below which $G$ is negative. Hence
the same is true for $g_{\varphi\varphi}$. Thus $r_1$ (which is the
root of $g_{\varphi\varphi}$) is the place where CTCs ``appear", and
we will be interested in seeing what the light cones do as we move
towards $r_1$ from greater values of $r$.

\begin{figure}[!h]
\begin{center} \bigskip
\psfrag*{l2}[b][b]{\rd{$\ell_2$}} \psfrag*{l1}[b][b]{\rd{$\ell_1$}}
\psfrag*{c}[b][b]{$c$} \psfrag*{d}[r][r]{$d$}
\psfrag*{d1}[r][r]{$d$} \psfrag*{1}[r][r]{$1$}
\psfrag*{rp}[l][l]{\gr{$r_+$}} \psfrag*{rm}[l][l]{\gr{$r_-$}}
\psfrag*{M}[l][l]{$M$} \psfrag*{t}[l][l]{\bl{$\partial_t$}}
\psfrag*{fi}[l][l]{\bl{$\partial_{\varphi}$}}
\psfrag*{ra}[c][l]{\gr{$\le a^{-1}$}}
\includegraphics[keepaspectratio, width=\textwidth]{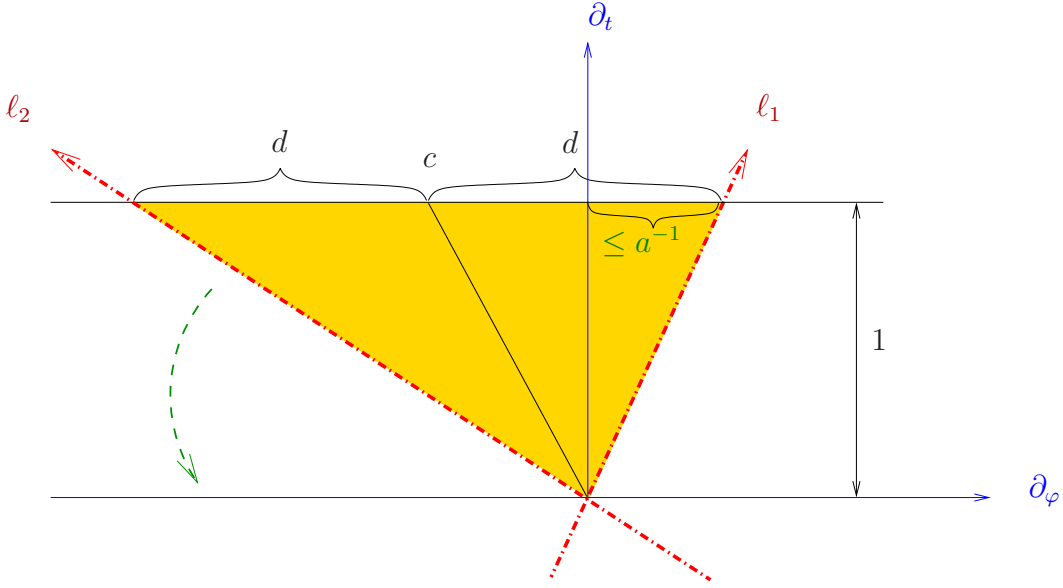}
\end{center}
\caption{\label{cone-fig} The light cone in the $t\varphi$-plane,
between $r_0$ and $r_1$. We draw the light cones with respect to
the Killing vector fields $\partial_t,\partial_{\varphi}$. As we
move towards the time machine region, i.e.\ towards $r_1$, $\ell_2$
lowers to horizontal while $\ell_1$ is less than or equal $a^{-1}$.}
\end{figure}

In order to see how the light cones tilt and widen in the
$t\varphi$-planes, we want to know for what values $y$ is the
direction $v=\partial_t+y\partial_{\varphi}$ lightlike, i.e.\ for
which $y$ we have $g(v,v)=0$. These values are the solutions of the
equation
\begin{description}
\item[]
$y^2g_{\varphi\varphi} + 2yg_{t\varphi} + g_{tt} = 0$,\qquad hence
\item[]
$y = \left(-g_{t\varphi}\pm \sqrt{g_{t\varphi}^2 -
g_{\varphi\varphi}g_{tt}}\right)/g_{\varphi\varphi}$.\quad
%
\comment{More precisely, we are interested in the values $ry$
because $\varphi$ is a spherical coordinate.} Let
\item[]
$c:=(-g_{t\varphi})/g_{\varphi\varphi}$,\quad
for \underbar{c}enter of light cone,
\item[]
$d:=(\sqrt{g_{t\varphi}^2 -
g_{\varphi\varphi}g_{tt}})/g_{\varphi\varphi}$,\quad for
half-\underbar{d}iameter of light cone, see
Figure~\ref{cone-fig}.
\end{description}

The light cone is erect, i.e.\ is not tilted, where $c=0$. This is
where $E=0$, i.e.\ at $r_0:=e^2/2M$. At $r=r_0$ we have $E=0$, thus
$g_{tt}=-1$, $g_{t\varphi}=0$, and $g_{\varphi\varphi}=r_0^2+a^2$.
This is almost the same as in Minkowski spacetime, the only
difference being that $g_{\varphi\varphi}$ is not $r^2$, but a bit
larger. Hence the light cones at $r_0$ are ``standing", but they are
a bit narrower than at infinity (the bigger $a$ is, the narrower the
light cones). See also Figure~\ref{series-fig}.

We show that $r_0>r_1$. Since the function
$G(r)=r^2g_{\varphi\varphi}$ decreases monotonically as we approach
$0$ and $G(r_1)=0$, in order to show $r_0>r_1$ it is enough to show
$G(r_0)>0$, and in order to show this it is enough to show
$g_{\varphi\varphi}(r_0)>0$ (since $r_0\ne 0$). That this is the
case can easily be seen:
\begin{description}
\item[]
$g_{\varphi\varphi}(r_0)=r_0^2+a^2+a^2E=r_0^2+a^2=(e^4/4M^2)+a^2>0$\quad
since\quad $a\ne 0$.
\end{description}

Let's have a look at the roots of $g_{tt}$. The smaller root of
$g_{tt}$, i.e.\ $R_-=M-\sqrt{M^2-e^2}$, is bigger than $r_0$,
because of the following:
\begin{description}
\item[]
$\frac{e^2}{2M} < M-\sqrt{M^2-e^2}$\qquad iff
\item[]
$\frac{e^2}{2M}(M+\sqrt{M^2-e^2}) <
(M-\sqrt{M^2-e^2})(M+\sqrt{M^2-e^2})$\qquad iff
\item[]
$\frac{e^2}{2M}(M+\sqrt{M^2-e^2}) < e^2$\qquad iff
\item[]
$M+\sqrt{M^2-e^2}< 2M$,\quad which always holds for $e\ne 0$.
\end{description}

The ergosphere is between $R_+$ and $R_-$. With $R_->r_0>r_1$ we
thus have shown that the light cones are erect somewhere between the
ergosphere and the ``Time Machine'' (as we claimed in
section~\ref{expl-sec}). We note that the two event horizons are at
$r_{\pm}=M\pm\sqrt{M^2-e^2-a^2}$ (these are the roots of
$\Delta=r^2-2Mr+e^2+a^2$). For $r_+, r_-$ and $r_0,r_1$ see also
Figure~\ref{series-fig}. We have established the following, see
Figure~\ref{functions-fig}:
\[ R_+>r_+>M>r_->R_->r_0>r_1>0\ .\]

We are interested in the behavior of the light cones as we move from
$r_0$ towards $r_1$. We have just shown that in this interval the
time-axis is always timelike, i.e.\ that it is within the light
cone. We will now establish the following statements concerning the
light cones as depicted in Figure~\ref{cone-fig} as we move from
$r_0$ to $r_1$:
\begin{description}
\item[(1)]
the center $c$ of the light cone moves from zero to minus infinity,
monotonically,
\item[(2)]
the diameter of the light cone grows to infinity, monotonically,
\item[(3)]
the right side of the light cone is always ``slower" than $a^{-1}$.
\end{description}
As a consequence of (1) and (2), the left side of the light cone
lowers towards the horizontal plain. We could summarize these
statements as ``the light cone tilts in the negative $\varphi$
direction and opens up".

To prove (1) and (2), first we show that
\begin{description}
\item[]
$g_{\varphi\varphi}=r^2+a^2 + a^2(2Mr-e^2)/r^2$
\end{description}
decreases monotonically as we move from $r_0$ towards $r_1$. Its
radial derivative is
\begin{description}
\item[]
$(g_{\varphi\varphi})' = 2[r+a^2(e^2-Mr)/r^3]$.
\end{description}
By $0<r\le r_0=\frac{e^2}{2M}<\frac{e^2}{M}$ we have $(g_{\varphi\varphi})'\ge 0$.
Thus, $g_{\varphi\varphi}$ decreases monotonically as we move from $r_0$ to $r_1$. Furthermore,
the reader is reminded that $g_{\varphi\varphi}$ is
positive in the open interval $]r_1,r_0[$ and zero for $r_1$.

Now, as can be seen in Figure~\ref{functions-fig}, $-g_{t\varphi}$
is negative and decreases monotonically as we move from $r_0$ to
$r_1$. Hence $c=-g_{t\varphi}/g_{\varphi\varphi}$
is negative and approaches monotonically minus infinity as we move
from $r_0$ to $r_1$. This proves (1).

To prove (2), we first compute:
\begin{description}
\item[]
$g_{t\varphi}^2-g_{\varphi\varphi}g_{tt}=a^2E^2-(r^2+a^2+a^2E)(E-1)=r^2+a^2-Er^2$
\item[]
$= r^2+a^2+e^2-2Mr$.
\end{description}
This is always positive for $r\le r_0$, and since it is a
second-order polynomial in $r$ whose roots are larger than $r_0$, it
increases as we move from $r_0$ to $r_1$. The same is true for the
square root of the above expression. Since $g_{\varphi\varphi}$ decreases
monotonically, this shows that the ``diameter" of the light cone
increases monotonically as we move from $r_0$ towards $r_1$. It
approaches infinity because $r_1$ is the root of
$g_{\varphi\varphi}$. Thus, (2) has been proved.

It remains to show that $c+d \le a^{-1}$, i.e.\ $y_1\le a^{-1}$ where\\
$y_1 :=\left(-g_{t\varphi} + \sqrt{g_{t\varphi}^2 -
g_{\varphi\varphi}g_{tt}}\right)/g_{\varphi\varphi}$. Let us use the
notation $S:=\sqrt{g_{t\varphi}^2 - g_{\varphi\varphi}g_{tt}}$.  Now
\begin{description}
\item[]
$y_1 =(-g_{t\varphi} + S)/g_{\varphi\varphi} =
(S-g_{t\varphi})(S+g_{t\varphi})/(g_{\varphi\varphi}(S+g_{t\varphi}))=$
\item[]
$[(g_{t\varphi}^2-g_{\varphi\varphi}g_{tt})-g_{t\varphi}^2]/[g_{\varphi\varphi}(S+g_{t\varphi})]=$
\item[]
$-g_{tt}/(S+g_{t\varphi})$.
\end{description}

We are interested in the behavior of the above expression as we move
$r$ from $r_0$ towards $r_1$. When substituting the values for
$g_{tt}$ and $g_{\varphi\varphi}$ we get
\begin{description}
\item[]
$y_1=(r^2-2Mr+e^2)/[r^2S+ae^2-2Mar]\le$
\item[]
$(r^2-2Mr+e^2)/(ar^2-2Mar+ae^2) = a^{-1}$
\end{description}
because $S=\sqrt{r^2+a^2+e^2-2Mr}\ge a$ if $r\le r_0$. With this, (3) has
been proved.

\section{A generic phenomenon?}\label{generic-sec}

We have seen in sections~\ref{expl-sec} and \ref{app-sec} that if
you want to time travel in a Kerr-Newman spacetime, you have to
orbit around the $\theta=0$ axis in the direction opposite to the
rotation carried out by the singularity. In other words, a time
traveler must counter-rotate with the singularity. The same happens
in Kerr spacetime, see section~\ref{kn-sec}. As it turns, however,
the phenomenon of counter-rotation is not limited to these important
classes of spacetimes.

The same counter-rotation phenomenon is present in van Stockum's
rotating dust solution (\cite{Stoc}), in the Tipler - van Stockum
fast-rotating cylinder (\cite{Tip74}, \cite{Tip77}), and in
G\"odel's rotating universe (\cite{God}). Counter-rotation in the first two examples is
discussed in \cite{ANW}, for counter-rotation in the third
example see \cite{AMNGodelspt}. Moreover, the same phenomenon is
also present in the case of  Gott's cosmic strings based CTCs, cf.\
\cite{Gott}, \cite[Fig.~14, p.~108]{Gottbook}. There the CTCs
counter-rotate with the system formed by the pair of strings.

In sum, thus, in at least five of the most prominent examples of
spacetimes involving CTCs, the future direction on the CTCs opposes
the rotational sense of the source of the gravitational
field.\footnote{In case of asymptotically flat spacetimes
(like Kerr-Newman ones) the counter-rotating effect can be
formulated in an invariant way as we did in section~\ref{kerr-sec}
in this paper. For the rest of our examples, we use slightly
different invariant formulations. For example, in G\"odel's
spacetime an invariant formulation can be obtained by following
G\"odel's wording in \cite[p.271, lines 1-10]{God}. G\"odel uses
gyroscopes as ``compasses of inertia" in the sense of \cite{MTW} and
Rindler~\cite[p.197, under the name ``gyrocompass"]{Rind}.}

This raises a question: Is this counter-rotation an accident or is
it a mathematical or at least physical necessity in
some sense, e.g.\ under some suitable physical assumptions, all of
which must of course be satisfied by the five examples above? One of
these assumptions is that the CTCs in question are ``created'' by
rotation of matter, i.e.\ rotation of the gravitational source. In
its present form this question is somewhat loose and rather
programmatic. Let us thus give a more tangible formulation of this
question.
\bigskip

\goodbreak
\noindent {\bf Question 2.} In general, how important is it for the
CTCs to counter-rotate against the rotational sense of the
gravitating matter which brings about the CTCs? In particular, is
there any example of a spacetime where the CTCs are generated by
rotating matter and there is no counter-rotation effect?
\bigskip

We believe that our calculations and our arguments above and in
\cite{ANW} not only admit the relevance and interest in asking this
question, but moreover, strongly suggest that there might well be a
general principle at work: a principle which states that CTCs
generated by rotating matter must spiral in the opposite direction
from the rotational sense of the matter. We are not sure how much
work such a principle would offer toward a general understanding of
the mechanisms which bring about spacetime structures that contain
CTCs. But it strikes us as an observation which is potentially
crucial for such an endeavor, particularly because the phenomenon
seems so pervasive in such an important set of spacetimes with CTCs.

\noindent {\bf Acknowledgement:} We would like to express our thanks
to John Earman for helpful discussions and encouragement. We would
like to express our gratitude for very helpful discussions at
several points in the past to Attila Andai, G\'abor Etesi, Judit X.\
Madar\'asz, Endre Szab\'o  and L\'aszl\'o B.\ Szab\'o. Very special
thanks go to Ren\'ata Tordai for various help, e.g.\ careful
reading. We would like to thank two anonymous referees for
helpful suggestions. C.W.\ thanks the Department of Philosophy at
the University of Bern for its hospitality.  This research was
supported by the Hungarian National Foundation for scientific research grant
No.\ T43242.

\bigskip

\goodbreak
\noindent
Hajnal Andr\'eka and Istv\'an N\'emeti\\
andreka@renyi.hu\quad and\quad nemeti@renyi.hu\\
R\'enyi Institute of Mathematics\\
Budapest, 1364 Hungary\\
\bigskip

\noindent
Christian W\"uthrich\\
wuthrich@ucsd.edu\\
Department of Philosophy\\
University of California at San Diego\\
9500 Gilman Drive\\
La Jolla, CA 92093-0119, USA

\end{document}